\documentclass[sigconf,screen=true]{acmart}
\usepackage{url}

\usepackage[utf8]{inputenc}
\usepackage[T1]{fontenc}
\usepackage[font=small]{caption}
\usepackage{amsmath}
\usepackage{booktabs}
\usepackage{color}
\usepackage{subfig}
\usepackage{float}
\usepackage{balance} 

\newcommand{\ts}[1]{{#1}}
\newcommand{\hk}[1]{{#1}}
\newcommand{\new}[1]{{#1}}

\newcommand{\edit}[1]{{#1}}

\author{Taras Kucherenko}
\affiliation{\institution{KTH, Stockholm, Sweden}}
\email{tarask@kth.se}
\author{Patrik Jonell}
\affiliation{\institution{KTH, Stockholm, Sweden}}
\email{pjjonell@kth.se}

\author{Sanne van Waveren}
\affiliation{\institution{KTH, Stockholm, Sweden}}
\email{sannevw@kth.se}

\author{Gustav Eje Henter}
\affiliation{\institution{KTH, Stockholm, Sweden}}
\email{ghe@kth.se}

\author{Simon Alexanderson}
\affiliation{\institution{KTH, Stockholm, Sweden}}
\email{simonal@kth.se}

\author{ Iolanda Leite}
\affiliation{\institution{KTH, Stockholm, Sweden}}
\email{iolanda@kth.se}

\author{Hedvig Kjellstr{\"o}m}
\affiliation{\institution{KTH, Stockholm, Sweden}}
\email{hedvig@kth.se}

\renewcommand{\shortauthors}{Kucherenko et al.}

\copyrightyear{2020}
\acmYear{2020}
\setcopyright{acmcopyright}\acmConference[ICMI '20]{Proceedings of the 2020 International Conference on Multimodal Interaction}{October 25--29, 2020}{Virtual event, Netherlands}
\acmBooktitle{Proceedings of the 2020 International Conference on Multimodal Interaction (ICMI '20), October 25--29, 2020, Virtual event, Netherlands}
\acmPrice{15.00}
\acmDOI{10.1145/3382507.3418815}
\acmISBN{978-1-4503-7581-8/20/10}

\begin{document}
\fancyhead{} 

\title[Gesticulator: A framework for semantically-aware speech-driven gesture generation]{Gesticulator: A framework for semantically-aware\texorpdfstring{\\}{ }speech-driven gesture generation} 

\begin{abstract}  

During speech, people spontaneously gesticulate, which plays a key role in conveying information. Similarly, realistic co-speech gestures are crucial to enable natural and smooth interactions with social agents. Current end-to-end co-speech gesture generation systems use a single modality for representing speech: either audio or text. These systems are therefore confined to producing either acoustically-linked beat gestures or semantically-linked gesticulation (e.g., raising a hand when saying ``high''): they cannot appropriately learn to generate both gesture types.
We present a model designed to produce arbitrary
beat and semantic gestures together. Our deep-learning based model takes both acoustic and semantic representations of speech as input, and generates gestures as a sequence of joint angle rotations as output. 
The resulting gestures can be applied to both virtual agents and humanoid robots.
Subjective and objective evaluations confirm the success of our approach. The code and video are available at the project page \href{https://svito-zar.github.io/gesticulator/}{svito-zar.github.io/gesticulator}.


\end{abstract}


\settopmatter{printfolios=true} 
\maketitle


\section{Introduction}
\label{sec:intro}

When speaking, people often spontaneously produce hand gestures, also referred to as co-speech gestures. These co-speech gestures can accompany the content of the speech -- what is being said -- on all levels, from partial word meanings to situation descriptions~\cite{kopp2007speech}. 
\edit{Gesture generation is hence an important part of animation
, as well as of human-agent interaction research and applications.} 


Virtual agents have been developed for a diverse set of applications, such as serious gaming~\cite{mascarenhas2018virtual}, interpersonal skills training~\cite{swartout2006toward, monahan2018autonomous} or therapy systems ~\cite{ring2016real}. Interactions with these virtual agents have shown to be more engaging when the agent's verbal behavior is accompanied by appropriate nonverbal 
behavior~\cite{salem2011friendly}. Moreover, it has been shown that  manipulating gesture properties 
can influence user perception of an agent's emotions \cite{castillo2019we}.


\begin{figure}
\includegraphics[width=0.75\linewidth]{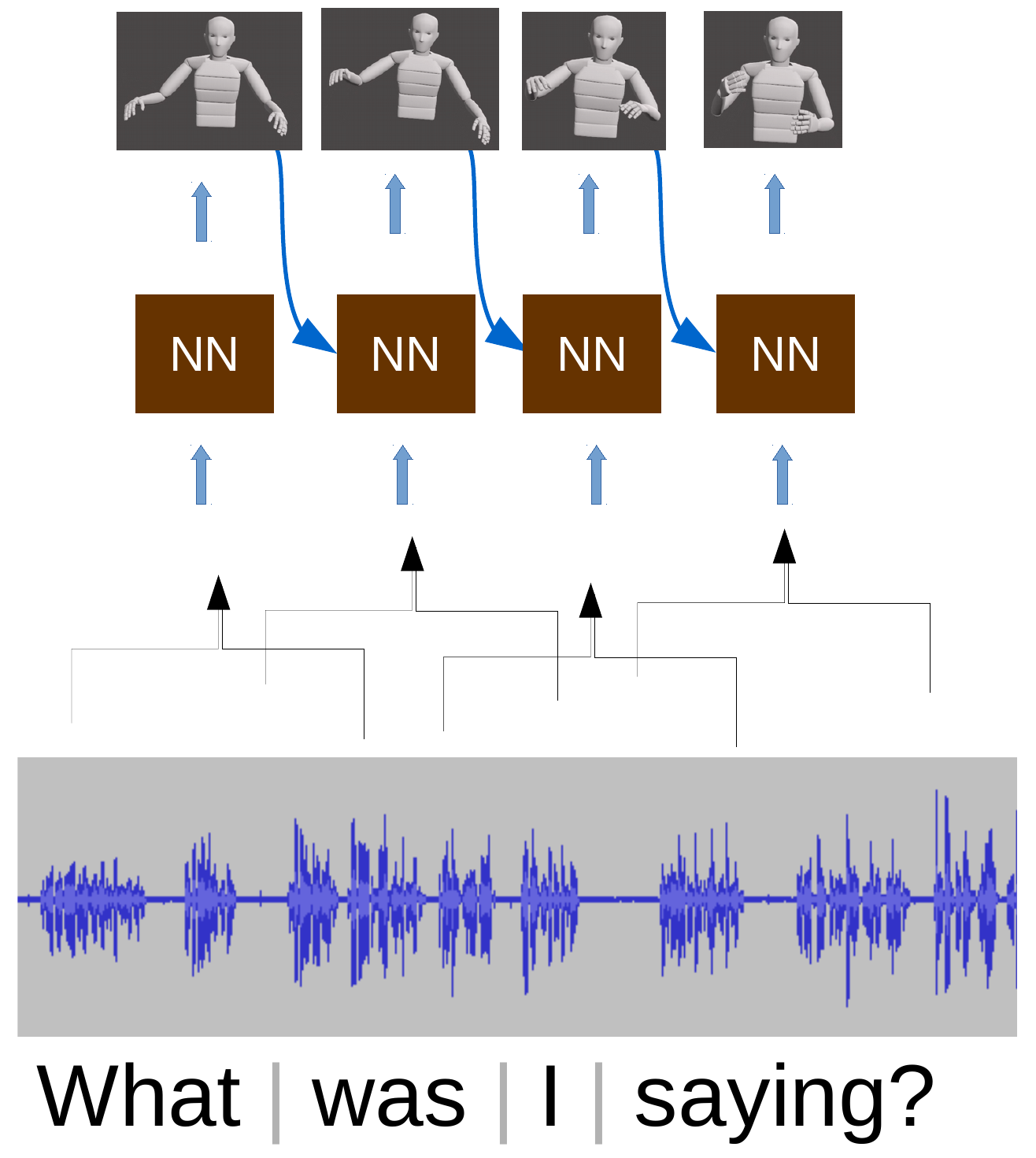}
\caption{Overview of the \new{proposed} autoregressive model.}
\label{fig:model_overview}
\vspace{-6mm}
\end{figure}

Traditionally, gesture generation for virtual agents has been done by various rule-based systems ~\cite{cassell2001beat,huang2012robot,salvi2009synface}. \edit{Those approaches are constrained by the discrete set of gestures they can produce.} Alongside recent advances in deep learning, data-driven approaches have increasingly gained interest for gesture generation ~\cite{kucherenko2018data, yoon2018robots, ahuja2019react}. While early work has considered gesture generation as a classification task which aims to deduce a specified gesture class~\cite{neff2008gesture,chiu2015predicting}, more recent work has considered it as a regression task which aims to produce continuous motion \cite{ yoon2018robots, alexanderson2020style}. We focus on the latter task: \emph{continuous gesture generation}. To date, prior work on continuous gesture generation has used a single input modality: either acoustic or semantic. In contrast, our work makes use of both these modalities to allow for semantic-aware speech-driven continuous gesture generation. The contributions of this work are the following:




\begin{enumerate}
    \item the first data-driven model that maps speech acoustic and semantic features into continuous 3D gestures; 
    \item a comparison contrasting the effects of different architectures and important modelling choices; 
    \item objective and subjective evaluations of the effect of the two speech modalities -- audio and semantics -- on the resulting gestures.
\end{enumerate}

We additionally extend a publicly available corpus of 3D co-speech gestures, the Trinity College dataset \cite{ferstl2018investigating}, with manual text transcriptions.
Video samples from our evaluations are provided at \href{https://vimeo.com/showcase/6737868}{vimeo.com/showcase/6737868}. 
\section{Background and Related Work}

\subsection{Background}
\label{sec:background}
While there are several theories on how gestures are produced by humans~\cite{mcneill1992hand,chu2016co,bickmore2004unspoken}, there is a consensus that speech and gestures correlate strongly~\cite{iverson1999hand,loehr2012temporal,pouwQuantifying,graziano2019referential}. In this section, we review some concepts relevant to our work, namely gesture classification, the temporal alignment between gestures and speech \new{as well as} the gesture-generation problem formulation.

\subsubsection{Co-Speech Gesture Types}
\label{subs:gs_types}
Our work is informed by the gesture classification by \mbox{McNeill} \shortcite{mcneill1992hand}, who distinguished the following gesture types: 
\begin{enumerate}
    \item \textit{Iconic} gestures represent some aspect of the scene;
    \item \textit{Metaphoric} gestures represent an abstract concept;
    \item \textit{Deictic} gestures point to an object or orientation;
    \item \textit{Beat} gestures are used for emphasis and usually correlate with the speech prosody (e.g., intonation and loudness).
\end{enumerate}
The first three gesture types\edit{, also called \textit{representational gestures},} depend on the content of the speech -- its semantics -- while the last type \new{instead} depends on the audio signal -- the acoustics. Hence, systems that ignore either aspect of speech can only learn to model a subset of human co-speech gesticulation.

\subsubsection{Gesture-Speech Alignment}
\label{subs:gs_alignment}
Gesture-speech alignment is an active research field covering several languages, including French \cite{ferre2010timing}, German \cite{bergmann2011relation}, and English \cite{loehr2012temporal,pouwQuantifying,graziano2019referential}. We focus on prior work on gesture-speech alignment for the English language.

In English, gestures typically lead the corresponding speech by, on average, 0.22 s (std 0.13 s) \cite{loehr2012temporal}; specifically, Pouw et al.~\cite{pouwQuantifying} aligned different gesture types with the peak pitch of the speech audio and found that the onset of beat gestures usually precedes the corresponding speech by 0.35 s (std 0.3), the onset of iconic gestures precedes speech by 0.45 s (std 0.4), and the onset of pointing gestures precedes speech by 0.38 s (std 0.4). 

Informed by these works, we take the widest range among the studies, plus some margin, for the time-span of the speech used to predict the corresponding gesture, and consider 1 s of future speech and 0.5 s of past speech as input to our model \new{detailed} in Sec.~\ref{sec:method}.

\subsubsection{\new{The} Gesture-Generation \new{Problem}}

We frame the problem of speech-driven gesture generation as follows: given a sequence of speech features $\boldsymbol{s} = [s_t]_{t=1:T}$ 
the task is to generate a corresponding pose sequence $\boldsymbol{\hat{g}} = [\hat{g}_t]_{t=1:T}$ of gestures that an agent might perform while uttering this speech. Here, $t=1:T$ \new{denotes} a sequence of vectors for $t$ in $1$ to $T$.

Each speech segment $\boldsymbol{s}_t$ is represented by several different features, such as acoustic features (e.g., spectrograms), semantic features (e.g., word embeddings) or a combination of the two.
The ground-truth pose $\boldsymbol{g}_t$ and the predicted pose $\boldsymbol{\hat{g}}_t$ at the same time instance $t$ can be represented in 3D space as a sequence of joint rotations:
$
\boldsymbol{g}_t = [\alpha_{i,t}, \beta_{i,t}, \gamma_{i,t}] _{i=1:n}\text{,}
$ 
$n$~being the number of keypoints of the body 
and $\alpha$, $\beta$ and $\gamma$ 
representing rotations in three axes.
\subsection{Related Work}
\label{sec:related}
\vspace{-1mm}
As this work contributes toward data-driven gesture generation, we confine our review to these methods. 

\subsubsection{\new{Audio-Driven Gesture Generation}}
Most prior work on data-driven gesture generation has used the audio-signal as the only speech-input modality in the model \new{ \cite{sadoughi2017speech, hasegawa2018evaluation, kucherenko2019analyzing, ginosar2019learning, ferstl2020adversarial}}. For example, Sadoughi and Busso~\shortcite{sadoughi2017speech} trained a probabilistic graphical model to generate \ts{a discrete set of} gestures based on the speech audio-signal, \new{using discourse functions as constraints}. 
Hasegawa et al.~\shortcite{hasegawa2018evaluation} developed a more general model capable of generating arbitrary 3D motion using a deep recurrent neural network, applying smoothing as postprocessing step. Kucherenko et al.~\shortcite{kucherenko2019analyzing} extended this work by applying representation learning to the human pose and reducing the need for smoothing. Recently, Ginosar et al.~\shortcite{ginosar2019learning} applied a convolutional neural network \new{ with adversarial training} to generate 2D poses from spectrogram features. However, driving either virtual avatars or humanoid robots requires 3D joint angles. Ferstl et al.~\cite{ferstl2020adversarial} followed the approach of adversarial training and applied it to a recurrent neural network together with a gesture phase classifier. \edit{Our model differs from these systems in that
it leverages both the audio signal and the text transcription for gesture generation.} 

\subsubsection{
\new{Text-Transcription-Driven Gesture Generation}%
}
Several recent works mapped from text transcripts to co-speech gestures. Ishi et al.~\shortcite{ishi2018speech} generated gestures from text input through a series of probabilistic functions: Words were mapped to word concepts using WordNet \cite{miller1995wordnet}, which then were mapped to a gesture function (e.g., iconic or beat), which in turn were mapped to clusters of 3D hand gestures. Yoon et al.~\shortcite{yoon2018robots} 
learned a mapping from the utterance text to gestures using a recurrent neural network. The produced gestures were aligned with audio in a post-processing step. 
Although these works capture important information from text transcriptions, they may fail to reflect the strong link between gestures and speech acoustics such as intonation, prosody, and loudness \cite{pouw2019gesture}. 


\subsubsection{
\new{Multimodal Gesture-Generation Models}%
}

Only a handful of works have used multiple modalities of the speech to predict matching gestures. The model in Neff et al.~\shortcite{neff2008gesture} predicted gestures based on text, theme, rheme, and utterance focus. They also incorporated text-to-concept mapping. Concepts were then mapped to a set of 28 discrete gestures in a speaker-dependent manner. Chiu et al~\shortcite{chiu2015predicting} used both audio signals and text transcripts as input, to predict a total of 12 gesture classes using deep learning. Our approach differs from these works, as we aim to generate a wider range of gestures: 
rather than predicting a discrete gesture class, our model 
produce\ts{s} arbitrary gestures as a sequence of 3D poses.


\subsubsection
{\new{Regarding Motion Continuity}}
\new{Separate from the input modalities of the system is the \hk{aspect of visual motion quality}.
Continuous gesture generation can avoid the concatenation-point discontinuities exhibited by playback-based approaches such as motion graphs \cite{arikan2002interactive,kovar2002motion}.
That said, comparatively few approaches to continuous gesture generation explicitly try to enforce continuity in the generated pose sequence.
Instead, they rely on postprocessing to increase smoothness as in \cite{hasegawa2018evaluation}.
Yoon et al.~\cite{yoon2018robots} include a velocity penalty in training that discourages jerky motion.
The recurrent connections used in several models \cite{hasegawa2018evaluation,ferstl2018investigating,yoon2018robots} can also act as a pose memory that may help the model to produce smooth output motion. Autoregressive motion models have recently demonstrated promising results in probabilistic audio-driven gesture generation \cite{alexanderson2020style}.  
In this paper, we \hk{similarly} investigate autoregressive connections for improving motion quality, which explicitly provide the most recent poses as input to the model when generating the next pose.

}


%
\section{Training and Test Data}
\label{sec:data}

We develop our gesture generation model using machine learning: we learn a gesture estimator $\boldsymbol{\hat{g}} = F(\boldsymbol{s})$ based on a dataset of human gesticulation, where we have both speech information $\boldsymbol{s}$ (acoustic and semantic) and gesture data $\boldsymbol{g}$. 
%
%
\ts{For this work, we specifically used} the Trinity Gesture Dataset~\cite{ferstl2018investigating}, comprising 244 minutes of audio and motion capture recordings of a male actor speaking freely on a variety of topics. We 
removed lower-body data, retaining 15 upper-body joints out of the original 69. 
Fingers were not modelled due to poor data quality.

To obtain semantic information for the speech, we first transcribed the audio recordings using Google Cloud automatic speech recognition (ASR), followed by thorough manual review to correct recognition errors and add punctuation \new{for both the training and test parts of the dataset}. 
\edit{The same data was used by the GENEA 2020 gesture generation challenge\footnote{\href{https://tinyurl.com/yxmdvxkq}{genea-workshop.github.io/2020/\#gesture-generation-challenge}} and has been made publicly available in the original dataset repository\footnote{\href{https://trinityspeechgesture.scss.tcd.ie/}{trinityspeechgesture.scss.tcd.ie}}.}

\subsection{Test-Segment Selection}
\label{subs:segm_select}
Two 10-minute recordings from the dataset were held out from training. We selected 50 segments of 10 s for testing: 30 random segments and 20 \emph{semantic segments}, in which speech and recorded gestures were semantically linked. Three human annotators marked time instants where the recorded gesture was semantically linked with the speech content. Instances where all three annotators agreed (\new{within} 5 s tolerance) were used as semantic segments in our experiments.


\subsection{Audio-Text Alignment}
\label{subsec:allign}
\begin{figure}
\begin{center}
\includegraphics[width=0.85\linewidth]{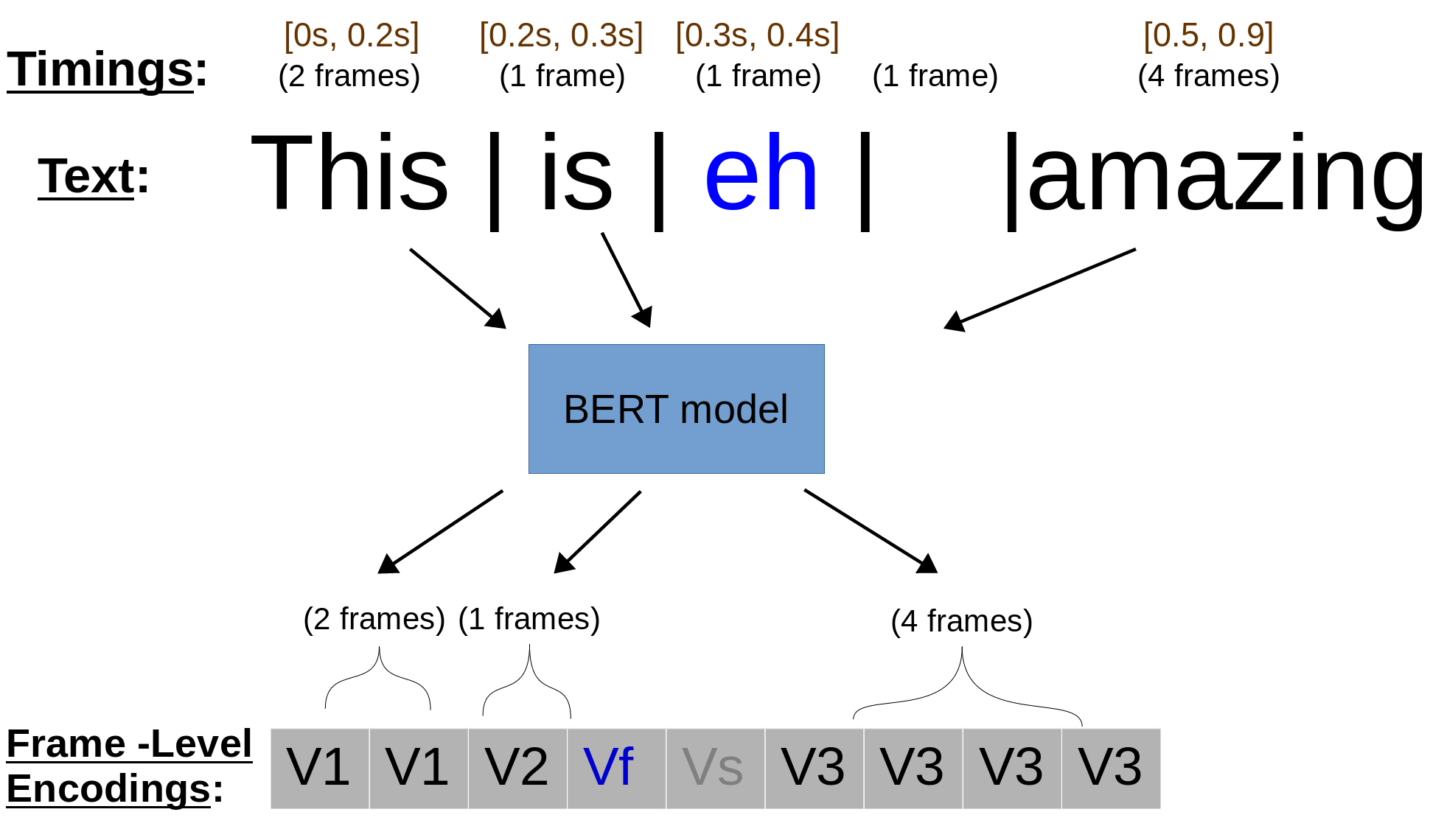}
\caption{Encoding text as frame-level features. First, the sentence (omitting filler words) is encoded by BERT~\cite{devlin2018bert}. We thereafter repeat each vector according to the duration of the corresponding word. Filler words and silence are encoded as fixed vectors, here denoted Vf and Vs.} 
\label{fig:text_enc}
\end{center}
\end{figure}
Text transcriptions and audio typically have different sequence lengths. 
To overcome this, we encode words into frame-level features as illustrated in Figure \ref{fig:text_enc}. First, the sentence, excluding filler words, is encoded by BERT~\cite{devlin2018bert}, which is the state-of-the-art model in natural language processing (NLP). 
We encode filler words and silence, which do not contain semantic information, as special, fixed vectors $V_f$ and $V_s$, respectively. Filler words typically indicate a thinking process and can occur with a variety of gestures. Therefore, we set the text feature vector $V_f$ during filler words equal to the average of the feature vectors for the most common filler words in the data. Silence typically has no gesticulation \cite{graziano2018speech}, so the silence feature vector $V_s$ was made distinct from all other encodings, by setting all elements equal to \textminus{}15. 
Finally, we use timings from the ASR system to nonuniformly upsample the text features, such that both text and audio feature sequences have the same length and timings.
\ts{This is a standard text-speech alignment method in the closely-related field of speech synthesis \cite{wu2016merlin}.}
\section{Speech-Driven Gesture Generation}
\label{sec:method}

This section describes our proposed method for generating upper-body motion from speech acoustics and semantics. 

\subsection{Feature Types}
We base our features on the state of the art in speech audio and text processing. 
Throughout our experiments, we 
use frame-synchronized features with 20 fps.

Like previous research in gesture generation \cite{ginosar2019learning,ferstl2018investigating}, we represent speech audio by log-power mel-spectrogram features. For this, we extracted 64-dimensional acoustic feature vectors using a window length of 0.1 s and hop length 0.05 s (giving 20 fps). 

For semantic features, we use BERT \cite{devlin2018bert} pretrained on English Wikipedia: each sentence of the transcription is encoded by BERT resulting in 768 features per word, aligned with the audio as described in Sec.~\ref{subsec:allign}. We supplement these by five frame-wise scalar features,
listed
in Table~\ref{tab:text_features}.

\begin{table}
  \caption{Text and duration features for each frame.}
  \label{tab:text_features}
\begin{tabular}{@{}l@{}}
    \toprule
    BERT encoding of \new{the} current word \\
    Time elapsed from the beginning of the word (in seconds)\\
    Time left until the end of the word (in seconds)\\
    Duration of this word (in seconds)\\
    Relative progress through the word (in \%) \\
    Speaking rate of this word (in syllables/second) \\
    \bottomrule
\end{tabular}
\end{table}

To extract motion features, the motion-capture data was downsampled to 20 fps and the joint angles were converted to an exponential map representation~\cite{grassia1998practical} relative to a T-pose; this is common in computer animation. \ts{We verified that the resulting features did not contain any discontinuities.}
Thereafter, we reduced the dimensionality by applying PCA and keeping 
92\% of the variance of the training data, similar to \cite{yoon2018robots}. This resulted in 12 components.

\subsection{Model Architecture and Training}

\begin{figure}
\begin{center}
\includegraphics[width=0.85\linewidth]{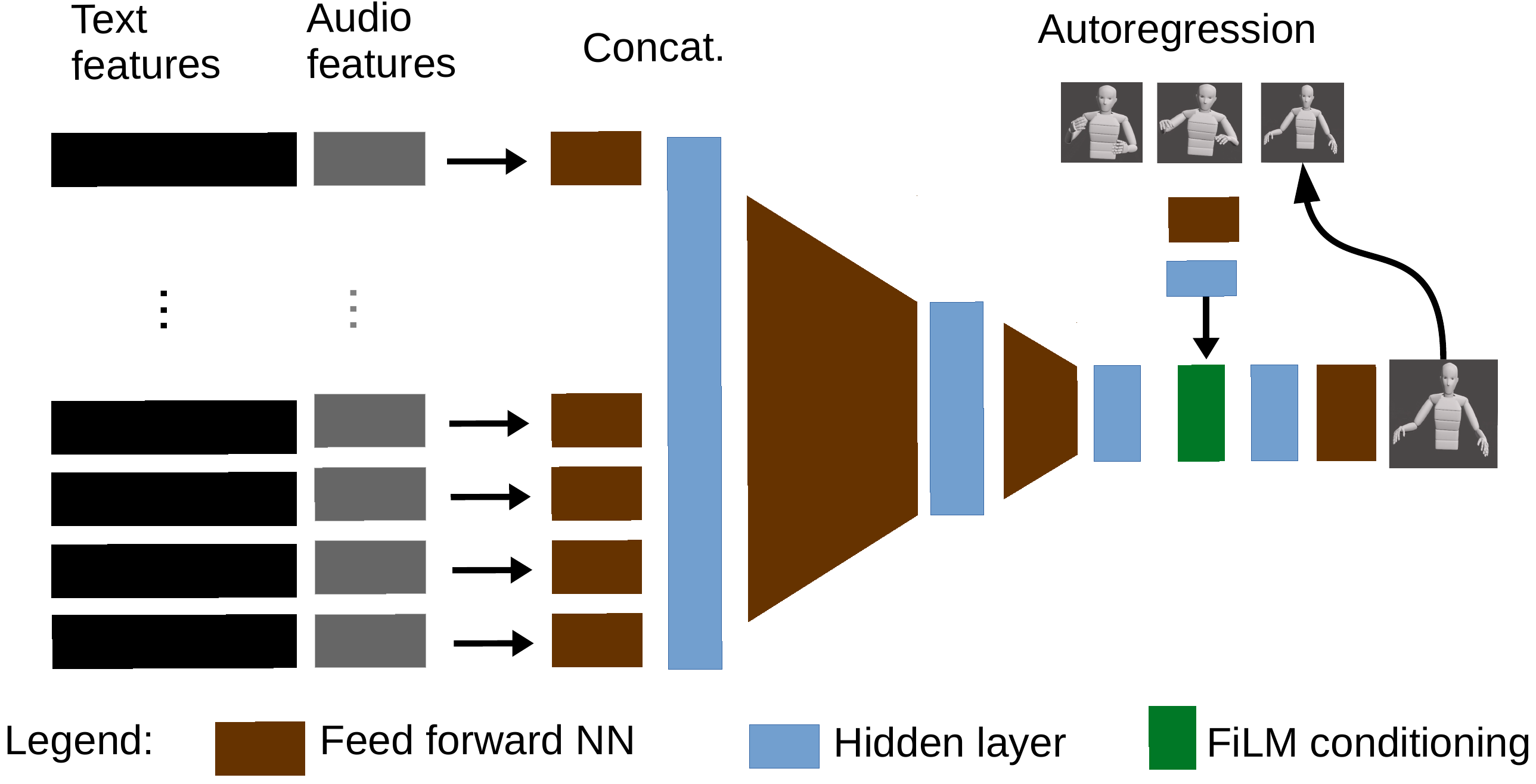}
\caption{Our model architecture. Text and audio features are encoded for each frame and the encodings concatenated. Then, several fully-connected layers are applied. The output pose is fed back into the model in an autoregressive fashion.}
\label{fig:model_arch}
\vspace{-4mm}
\end{center}
\end{figure}

\edit{We believe that a simple model architecture is preferable to a more complex one, everything else being equal. Hence, the intent of this work was to develop a straightforward model that solves the studied task.} 
Figure \ref{fig:model_arch} illustrates our model architecture. First, the text and audio features of each frame are jointly encoded by a feed-forward neural network to reduce dimensionality.
To provide more input context for predicting the current frame, we pass a sliding window spanning 0.5 s (10 frames) of past speech and 1 s (20 frames) of future speech 
features over the encoded feature vectors. These time spans are grounded in research on gesture-speech alignment\new{, as reviewed in Sec.~\ref{subs:gs_alignment}}. 
The encodings inside the context window are concatenated into a long vector and passed through several fully-connected layers. The model is also autoregressive: we feed preceding model predictions back to the model as can be seen in the figure, to ensure motion continuity. To condition on the information from the previous poses, we use FiLM conditioning \cite{perez2018film}, which generalizes regular concatenation. FiLM applies element-wise affine transforms $ \mathrm{FiLM} (\boldsymbol{x}, \boldsymbol{\alpha}, \boldsymbol{\beta}) = \boldsymbol{x} * \boldsymbol{\alpha} + \boldsymbol{\beta} $ to network activations $\boldsymbol{x}$, where scaling $\boldsymbol{\alpha}$ and offset $\boldsymbol{\beta}$ vectors are produced by a neural net taking other information (here previous poses) as input.
The final layer of the model and of the conditioning network for FiLM are linear to not restrict the attainable output range.

\subsection{Training Procedure}
We train our model on sequences of aligned speech audio, text, and gestures from the dataset. Each training sequence contains 70 consecutive frames from a larger recording. The first 10 and the last 20 frames establish context for the sliding window, while the 40 central frames are used for training. The model is optimized end-to-end for 100 epochs using stochastic gradient descent (SGD) and Adam \cite{kingma2014adam} to minimize the loss function $\mathrm{loss}(\boldsymbol{g},\boldsymbol{\hat{g}}) = \mathrm{MSE}(\boldsymbol{g}, \boldsymbol{\hat{g}} ) + \lambda ~ \mathrm{MSE}(\Delta \boldsymbol{g}, \Delta \boldsymbol{\hat{g}})$,
here $\boldsymbol{g}$ and $\Delta \boldsymbol{g}$ are the ground-truth position and velocity, $\boldsymbol{\hat{g}}$ and $\Delta \boldsymbol{\hat{g}}$ are the same quantities for the model prediction and MSE stands for Mean Squared Error. The weight $\lambda$ was set empirically to 0.6. Our velocity penalty can be seen as an improvement on the penalty used by Yoon et al.~\shortcite{yoon2018robots}. Instead of penalizing the absolute value of the velocity, we enforce velocity to be close to that of the ground truth.

During development, we observed that information from previous poses (the autoregression) tended to overpower the information from the speech: our initial model moved independently of speech input and quickly converged to a static pose. This is a common failure mode in generative sequence models, \new{cf.\ \cite{chen2017variational,henter2019moglow}}. To counteract this, we pretrain our model without autoregression for the first seven epochs (a number chosen empirically), before letting the model receive autoregressive input. This pretraining helps the network learn to extract useful features from the speech input, an ability which is not lost during further training. \new{Additionally, while full training begins without any teacher forcing (meaning that the model receives its own previous predictions as autoregressive input instead of the ground-truth poses), this is annealed over time: after one epoch, the model receives the ground-truth poses instead of its own prediction (for two consecutive frames) every 16 frames, which increased to every eight frames after another epoch, to every four frames after the next epoch, and then to every single frame after that. Hence, after five epochs of training with autoregression, our model has full teacher forcing: it always receives the ground-truth poses for autoregression.
This procedure greatly helps with learning a model that properly integrates non-autoregressive input.}

\subsection{Hyper-Parameter Settings}

For the experiments in this paper, we used the hyper-parameter search tool Tune \cite{liaw2018tune}. We performed random search over 600 configurations with velocity loss as the only criterion, obtaining the following hyper-parameters: Speech-encoding dimensionality 124 at each of 30 frames, producing 3720 elements after concatenation. 
The three subsequent layers had 612, 256, and 12 or 45 nodes (the output dimensionality with or without PCA). Three previous poses were encoded into a 512-dimensional conditioning vector. 
The activation function was $\tanh$, the batch size was 64 and the learning rate 10\textsuperscript{-4}. For regularization, we applied dropout with probability 0.2 to each layer, except for the pose encoding, which had dropout 0.8 to prevent the model from attending too much to past poses.


\section{Evaluation Measures}

\new{In this section we describe the objective and subjective measures we used in our experiments (Secs.~\ref{sec:ablation} and \ref{sec:baselining}).}

\subsection{Objective Measures}
\label{subs:obj_meas}
There is no consensus in the field about which objective measures should be used to evaluate the quality of generated gestures. 
As a step towards common evaluation measures for the gesture generation field, we primarily use metrics proposed by previous researchers. 
Specifically, we evaluated the average values of \ts{root-mean-square error (RMSE)}, acceleration and jerk \new{(rate of change of acceleration)}, and acceleration histograms of the produced motion, in line with Kucherenko et al.~\shortcite{kucherenko2019analyzing}. To obtain these statistics, the gestures were converted from joint angles to 3D \new{joint} positions.

The acceleration and jerk were averaged over all frames \new{for all 14 3D joints (except for the hips, which were fixed)}.
To investigate the motion statistics in more detail, we also computed velocity histograms of the generated motion and compared those against histograms derived from the ground-truth test data. We calculated the relative frequency of different velocity values over time-frames in all 50 test sequences, split into bins of width 1 cm/s. 


\subsection{Subjective Measures}
\label{subsec:subj_measures}
\new{To investigate human perception of the gestures we conducted several user studies that all followed the same protocol and procedure.}

\subsubsection{Experiment Design}
We assessed the perceived human-likeness of the virtual character's motion and how the motion related to the character's speech using measures adapted from recent co-speech gesture generation papers \cite{ginosar2019learning,yoon2018robots}. Specifically, we asked the questions ``In which video...'': (Q1) ``...are the character's movements most human-like?''
     (Q2) ``...do the character's movements most reflect what the character says?''
     (Q3) ``...do the character's movements most help to understand what the character says?''
     (Q4) ``...are the character's voice and movement more in sync?''

\new{We used attention checks to filter out inattentive participants.}
For four of the six attention checks, we picked a random video in the pair and heavily distorted either the audio (in the 2nd and 17th video pairs) or the video quality (in the 7th and 21st video pairs). Raters were asked to report any video pairs where they experienced audio or video issues, 
and were automatically excluded from the study upon failing any two of these four attention checks.
In addition, the 13th and 24th video pairs presented the same video (from the random pool) twice. Here an attentive rater should answer ``no difference''. 

\subsubsection{Experimental Procedure}

Participants were recruited on Amazon Mechanical Turk (AMT) and \new{assigned to one specific comparison of two systems}; they could complete the study only once, and were thus only exposed to one \new{system pair}. Each participant was asked to evaluate 26 same-speech video pairs on the four subjective measures: 10 pairs randomly sampled from a pool of 28 random segments, 10 from a pool of 20 semantic segments, and 6 attention checks \new{(see above)}. These video pairs were then randomly shuffled.

Every participant first completed a training phase to familiarize themselves with the task and interface. This training consisted of five items not included in the analysis, with video segments not present in the study, showing gestures of different quality.
Then, during the experiment, the videos in each pair were presented side by side in random order and could be replayed as many times as desired. For each pair, participants indicated which video they thought best corresponded to a given question (one of Q1 through Q4 above), or that they perceived both videos to be equal in regard to the question.
\section{Ablation Study}
\label{sec:ablation}

\new{In this section}, we evaluate the importance of various model components by individually ablating them, training seven different system variants including the full model (see Table~\ref{tab:systems}).
\new{Comparisons against other gesture-generation approaches are reported in Sec.~\ref{sec:baselining}.}

\subsection{Objective Evaluation}
\label{sec:obj}

\new{In this section we report objective metrics, as described in Sec. \ref{subs:obj_meas}.}

\subsubsection{Average Motion Statistics}

Table \ref{tab:avg_jerk} illustrates acceleration and jerk, as well as RMSE, averaged over 50 test samples for the ground truth 
and the different ablations of the proposed method. Ground-truth statistics are given as reference values for natural motion. \new{We focus our analysis on the \textit{jerk}, since it is commonly used to evaluate the smoothness of the motion: the lower the jerk the smoother the motion is \cite{morasso1981spatial, uno1989formation}.}

\begin{table}
  \caption{The seven system variants \new{in the ablation study}}
  \label{tab:systems}
  \begin{tabular}{@{}l|l@{}}
    \toprule
    System & Description \\
    \toprule
    Full model & The proposed method \\
    No PCA & No PCA is applied to output poses\\ 
    No Audio & Only text is used as input\\
    No Text & Only audio is used as input \\
    No FiLM & Concatenation instead of FiLM\\ 
    No Velocity loss & The velocity loss is removed\\ 
    No Autoregression & The previous poses are not used \\ 
\end{tabular}
\end{table}

We can observe that the proposed model \new{exhibits lower jerk} than the original motion. This is probably because our model is deterministic and hence produces gestures closer to the mean pose.
Not using PCA results in higher acceleration and jerk, and made the model statistics closer to the ground truth. Our intuition for this is that PCA reduced variability in the data, which resulted in over-smoothed motion. 
Removing either audio or text input \new{reduced the jerk even further}. This is probably because \new{these ablations provide} a weaker input signal to drive the model, making it gesticulate closer to the mean pose. Both FiLM conditioning and the velocity penalty seem to have little effect on the motion statistics and are likely not central to the model. That autoregression is a key aspect of our system is clear from this evaluation: without autoregression, the model loses continuity and generates motion with excessive jerk. 
RMSE appears to not be informative. 
This is expected since there are many plausible ways to gesticulate, so the minimum-expected-loss output gestures do not have to be close to our ground truth.

\begin{table}
  \caption{Objective evaluation of our systems: mean and standard deviation over 50 samples.}
  \label{tab:avg_jerk}
  \begin{tabular}{@{}lccc@{}}
    \toprule
    System & Accel.~\new{(cm/s\textsuperscript{2})} & Jerk \new{(cm/s\textsuperscript{3})} & \ts{RMSE \new{(cm)}}  \\
    \toprule
    Full model &  \hphantom{0}37.6 $\pm${~\hphantom{0}4.3} & \hphantom{0}830 $\pm${~\hphantom{0}89}  & 11.4 $\pm${~11.8} \\
    \midrule
    No PCA & \hphantom{0}63.8 $\pm${~\hphantom{0}8.3} & 1332 $\pm${~192} & 13.0 $\pm${~14.7}\\
    No Audio & \hphantom{0}26.9 $\pm${~\hphantom{0}3.9}  & \hphantom{0}480 $\pm${~\hphantom{0}67} & 11.3 $\pm${~11.7}\\
    No Text & \hphantom{0}27.0 $\pm${~\hphantom{0}1.9} & \hphantom{0}715 $\pm${~\hphantom{0}63} & 10.9 $\pm${~11.3} \\
    No FiLM & \hphantom{0}44.2 $\pm${~\hphantom{0}6.6} & \hphantom{0}931 $\pm${~181} & 11.0 $\pm${~11.5}\\
    No Velocity loss & \hphantom{0}36.4 $\pm${~\hphantom{0}4.1} & \hphantom{0}779 $\pm${~\hphantom{0}93} & 11.4 $\pm${~12.3}\\
    No Autoregression & 120.3 $\pm${~19.2} & 3890 $\pm${~637} & 11.2 $\pm${~12.0}\\
    \midrule
    Ground truth   & \textbf{144.7} $\pm${~36.6} & \textbf{2322}  $\pm${~538}  & \textbf{0}
\vspace{-1mm}
\end{tabular}
\end{table}

\begin{figure}[!t]
\centering
\subfloat[Comparing different architectures.]
{\includegraphics[width=0.9\linewidth]{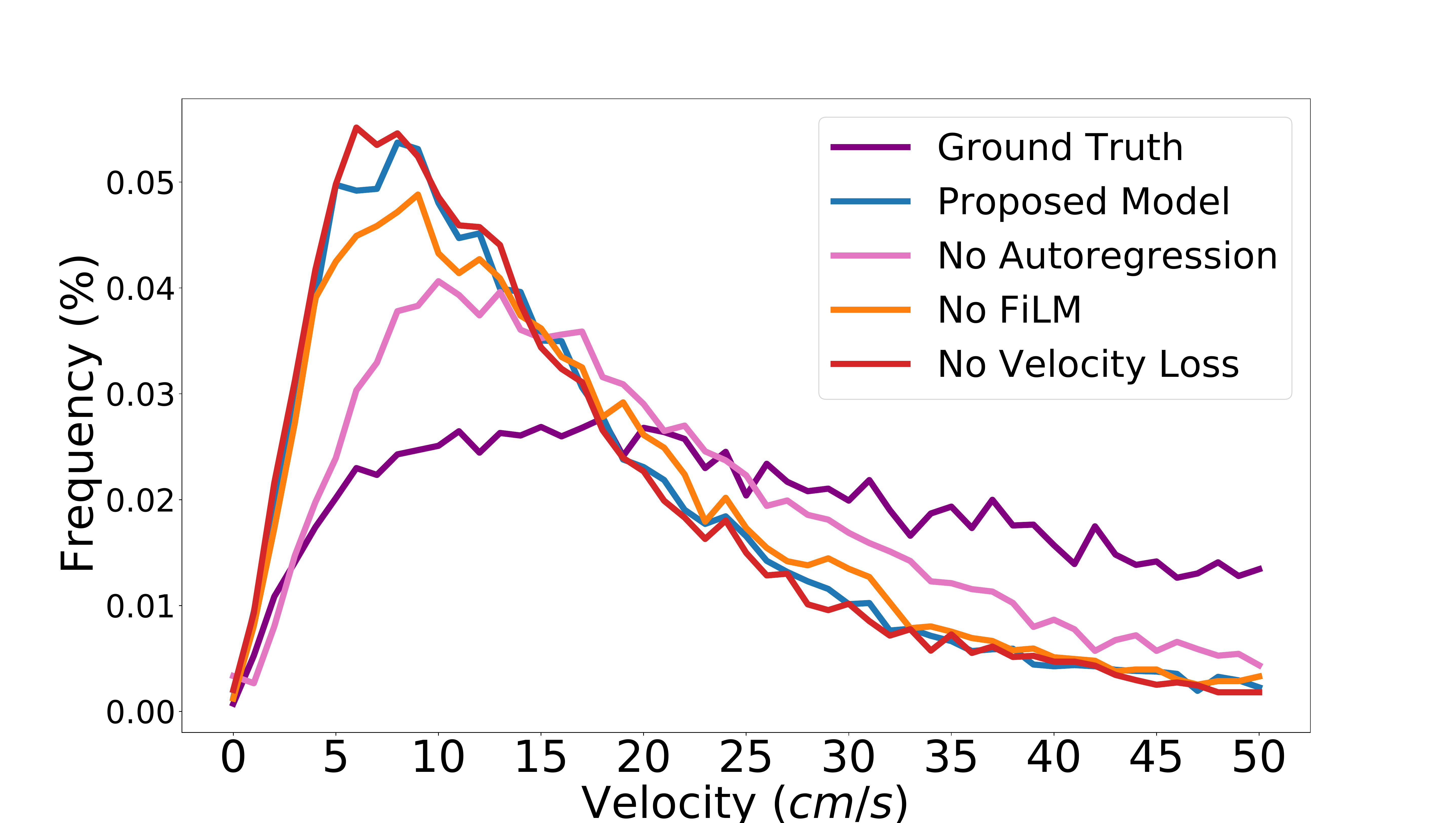}}
\label{fig:hist_arch}\\
\subfloat[Comparing different input/output data.]
{\includegraphics[width=0.9\linewidth]{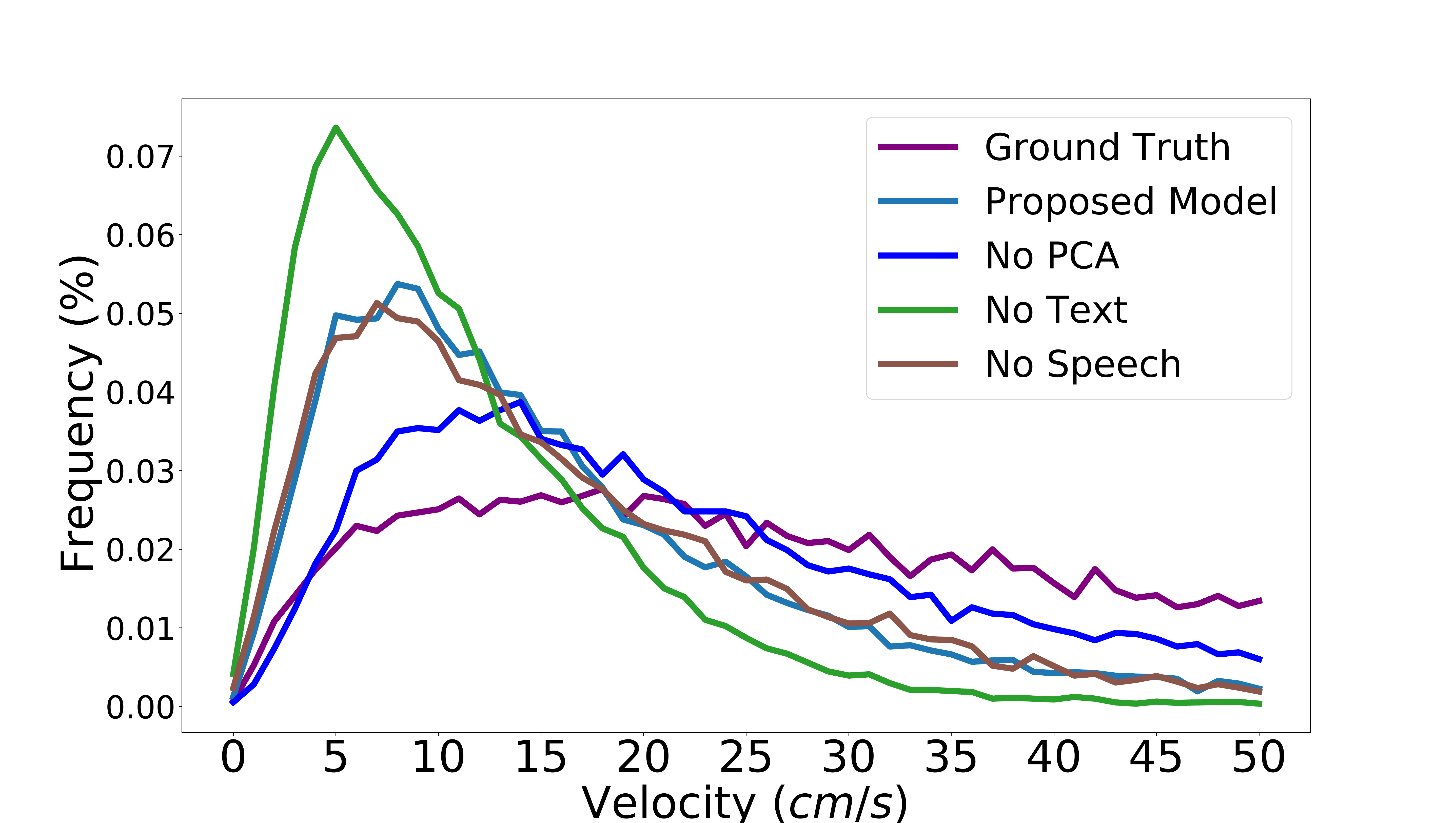}}
\label{fig:hist_data}
\caption{Velocity histograms \edit{of the wrist joints} for the ablation study.} 
\label{fig:num_eval}
\vspace{-4mm}
\end{figure}

\subsubsection{Motion \edit{Velocity} Histograms}
The values in Table \ref{tab:avg_jerk} were averaged over all time-frames and over all joints.
To investigate the motion statistics in more detail, we computed \edit{velocity} histograms of the generated motion and compared those against histograms derived from the ground-truth test data. \edit{As previous work has shown that wrist histograms are more informative than histograms averaged over all joints \cite{kucherenko2019analyzing}, we consider only left and right wrist joints.}

Figure \ref{fig:num_eval}a illustrates the \edit{velocity} histogram \edit{of the wrist joints} for the different model architectures and loss functions we considered. We observe 
two things: 1) the distributions are not influenced strongly by either FiLM conditioning or by velocity loss; and 2) \edit{autoregression reduces the amount of fast moments, making the velocity histogram more similar to the ground truth}.

Velocity histograms for different input/output data are shown in Figure~\ref{fig:num_eval}b. Removing PCA increases velocity, making the distribution more similar to the ground truth. In other words, training our model in the PCA space leads to reduced variability, which makes sense. We observe that 
excluding the text input makes the velocity smaller. This agrees with Table \ref{tab:avg_jerk} and probably means that without semantic information the model produces mainly beat gestures, whose characteristics differ from other gesture types. 
%
%
%
While these numerical evaluations are valuable, they say very little about people's perceptions of the generated gestures.

\subsection{First Perceptual Study}
\label{sec:subj}
To investigate human perception of the gestures we conducted several user studies.
This section reports on Perceptual Study 1, in which we evaluated participants' perception of a virtual character's gestures as produced by the seven variants of our model described in Table~\ref{tab:systems}. \new{The experimental procedure and evaluation measures (see Sec.~\ref{subsec:subj_measures}) were identical across all perceptual studies, including this one.} 
\new{Video samples from all systems in this study can be found at
\href{https://vimeo.com/showcase/6737868}{vimeo.com/showcase/6737868.}}



\begin{figure*}[ht]
\begin{center}
\includegraphics[width=0.95\textwidth]{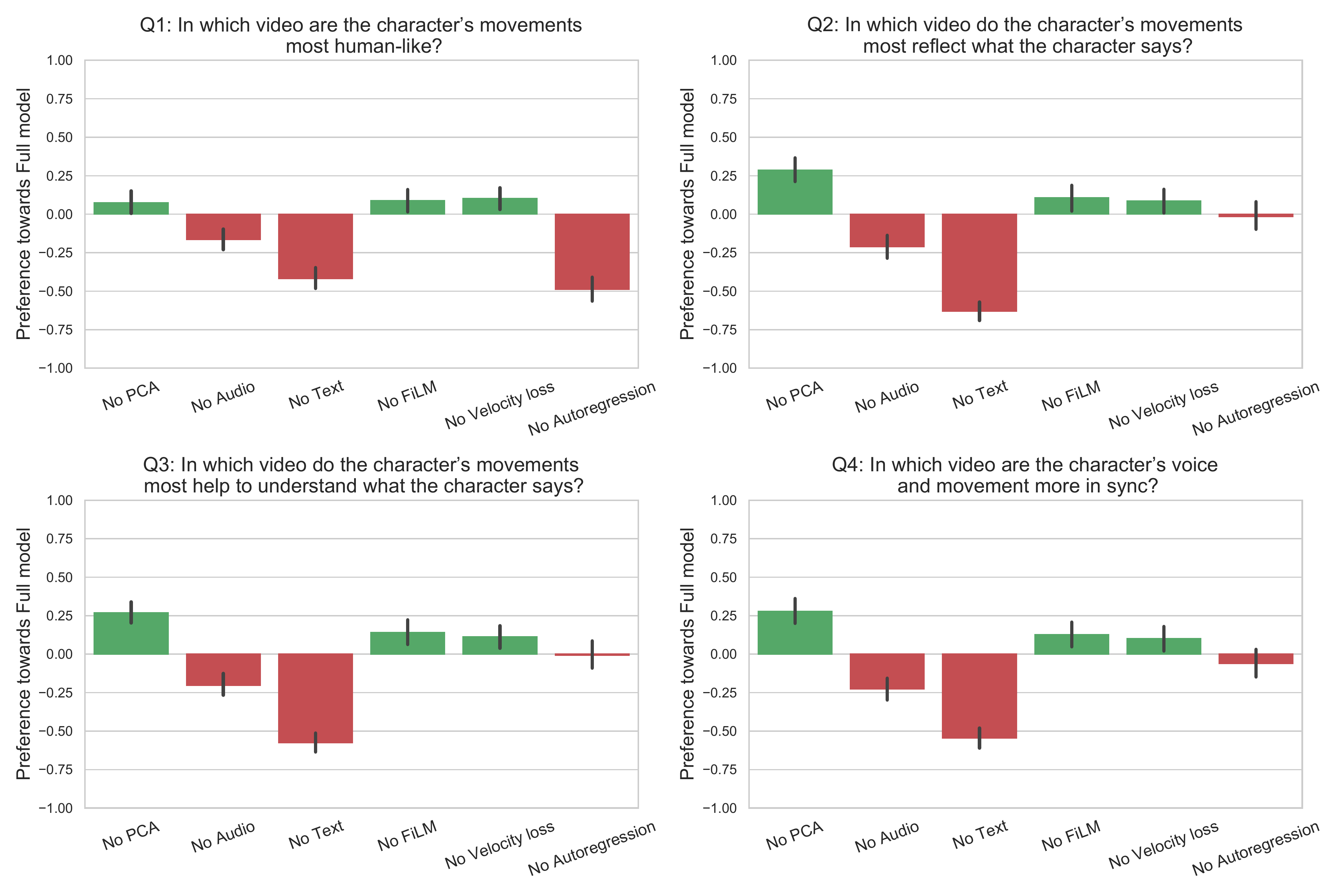}
\caption{Results of Perceptual Study 1: comparing different ablations of our model in pairwise preference tests. Four questions, listed above each bar chart, were asked about each pair of videos. The bars show the preference towards the full model (\new{higher values mean stronger preference}) with 95\% confidence intervals.}
\label{fig:all_models}
\vspace{-3mm}
\end{center}
\end{figure*}

In the comparison of system ablations (Perceptual Study 1), 123 participants ($\mu$ age = $41.8\pm12.3$; 52 male, 70 female, 1 other) remained after exclusion of 477 participants who failed the attention checks, experienced technical issues, or stopped the study prematurely. The majority were from the USA ($N$ = 120). 
\new{Each sub-study had between 19 and 21 participants.}
We conducted a binomial test excluding ties with Holm-Bonferroni correction of $p$-values to analyze the responses. 
(24 responses that participants flagged for technical issues were excluded.)
Our analysis was done in a double-blind fashion such that the conditions were obfuscated during analysis and only revealed to the authors after the statistical tests had been performed. 
The results are shown in Figure~\ref{fig:all_models}.

We can see from the evaluation of the ``No Text'' system that removing the semantic input drastically decreases both the perceived human-likeness of the produced gestures and how much they are linked to speech: participants preferred the full model over the one without text across all four questions asked with $p$\textless.0001.
This confirms that semantics are important for appropriate automatic gesture generation.

The ``No Audio'' model is unlikely to generate beats, and might not follow an appropriate speech rhythm 
. Results in Figure \ref{fig:all_models} confirm this: \new{participants preferred the full model over the one without audio across all four questions asked ($p$\textless.0001).}


Removing autoregression from the model only affected perceived naturalness, where it performed significantly worse ($p$\textless.0001), as shown in Figure \ref{fig:all_models}. This aligns with the findings from the objective evaluation: without autoregression the model produces jerky, unnatural-looking gestures, but the jerkiness does not influence whether gestures are semantically linked to the speech content.

\new{There was no statistical difference between the Full model and the model without FiLM conditioning in terms of Q1 and Q2, but the model without FiLM was preferred with $p$\textless.02 for Q3 and $p$\textless.04 for Q4. This suggests that FiLM conditioning was not helpful for the model and regular concatenation worked better.}

\new{Removing the velocity penalty did not have a statistical difference on user responses, except for reducing user preference on Q4 with $p$\textless.04, suggesting that this component is not critical for the model.} 


The model without PCA gave unexpected results. In videos, we see that removing PCA improved gesture variability. While for human-likeness, there was no statistical difference, ``No PCA'' was significantly better ($p$\textless.0001) on Q2, Q3 \new{and Q4} (see Figure \ref{fig:all_models}). 
In summary, participants preferred the system without PCA, so it was chosen as our final model for the remaining comparisons \new{(in Sec.~\ref{sec:baselining})}.

\subsection{Relation between objective and subjective evaluations}

\new{
Objective and subjective evaluations each have their pros and cons. 
In this subsection we analyze the empirical correlation between the two for the experiments reported here.

From, the ``No FiLM'' and ``No Velocity Loss'' conditions, we see that user ratings did not change much for ablations that only produced minor changes in motion statistics.
This is not surprising.
For models with low jerk compared to the ground truth, we can see that participants preferred the models where the jerk was closer to that of the ground truth motion (``No PCA'').
However, too high jerk was associated with unnatural motion (``No Autoregression''). These results seem to indicate that jerk analysis provides information about the human-likeness of the motion. 
}
\section{Additional Evaluations and Comparisons}
\label{sec:baselining}

\new{
The primary goal of this work is
to develop the first model for \emph{continuous gesture generation} that takes into account both the semantics and the acoustics of the speech. That said, we also benchmark our model against the state of the art in gesture generation. We compare the proposed approach to the model by Ginosar et al.~\cite{ginosar2019learning}, which is based on CNNs (convolutional neural networks) and GANs (generative adversarial networks), and therefore denoted CNN-GAN.


The hyper-parameters for the baseline method were fine-tuned by changing one parameter at a time and manually inspecting the visual quality of the resulting gestures on the validation dataset. The final hyper-parameters for the CNN-GAN \cite{ginosar2019learning} model were: batch size = 256, number of neurons in the hidden layer = 256, learning rate = 0.001, training duration = 300 epochs and $\lambda$ coefficient for the discriminator loss = 5.
This tuned system was compared against the best system (``No PCA'') identified in Sec.~\ref{sec:ablation}.


\subsection{\edit{Comparing with the state-of-the-art}}

\begin{table}
  \caption{Objective comparison of our systems with the state-of-the-art: mean and standard deviation over 50 samples}
  \label{tab:baseline_numerical}
  \begin{tabular}{@{}lccc@{}}
    \toprule
    System & Accel.~\new{(cm/s\textsuperscript{2})} & Jerk \new{(cm/s\textsuperscript{3})} \\
    \toprule
    Final model (no PCA) &  \hphantom{0}63.8 $\pm${~\hphantom{0}8.3} & 1330  $\pm${~192} \\
    CNN-GAN \cite{ginosar2019learning} & 254.7 $\pm${~31.8} &  5280  $\pm${~631}\\
    \midrule
     Ground truth  & 144.2 $\pm${~35.9} &  2315  $\pm${~530}
     
\end{tabular}
\end{table}


Like the previous experiments, we follow the objective evaluation setup described in Sec.~\ref{subs:obj_meas}.
Table \ref{tab:baseline_numerical} displays the average acceleration and jerk over 50 test sequences. We observe that the proposed method has acceleration and jerk values roughly half of those exhibited by the ground truth, while the CNN-GAN \cite{ginosar2019learning} baseline instead has twice the acceleration and jerk of the ground truth.

To investigate which model is preferred by human observers, we conducted another user study.
We evaluated participants' preference between the gestures 
as produced by the proposed models (No PCA) and CNN-GAN \cite{ginosar2019learning} (Perceptual Study 2). 
Video samples from this study can be found at
\href{https://vimeo.com/showcase/7127462}{vimeo.com/showcase/7127462}.

The study setup was the same described in Sec.~\ref{subsec:subj_measures}, except for three minor changes:
\begin{enumerate}
    \item We paid online participants more (5\$ instead of 3\$), since we realized that the effort required from participants in the previous study was higher than we had anticipated. 
    \item We clarified the instructions for reporting broken audio/video.
    \item We only asked questions Q1 (human-likeness of movements) and Q2 (movements reflect what the character says).
\end{enumerate}

In this study 27 participants ($\mu$ age = $41.7\pm11.3$; 14 male, 13 female) remained after exclusion of 43 participants based on the same criteria as before. 
The majority were from the USA ($N$ = 25).

\begin{figure}[!t]
\hfill
{\includegraphics[height=0.57\linewidth]{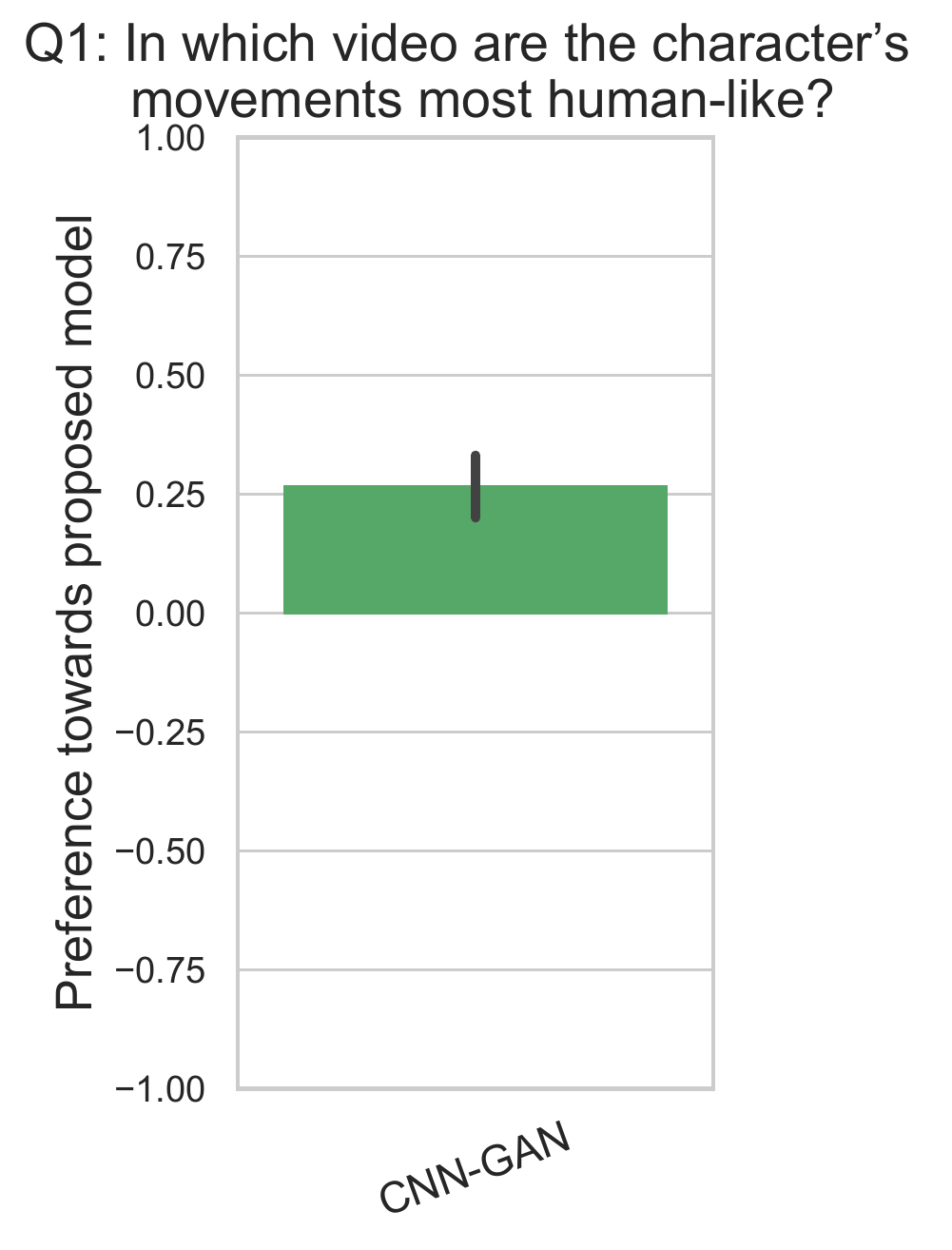}}
\hfill\hfill
{\includegraphics[height=0.57\linewidth]{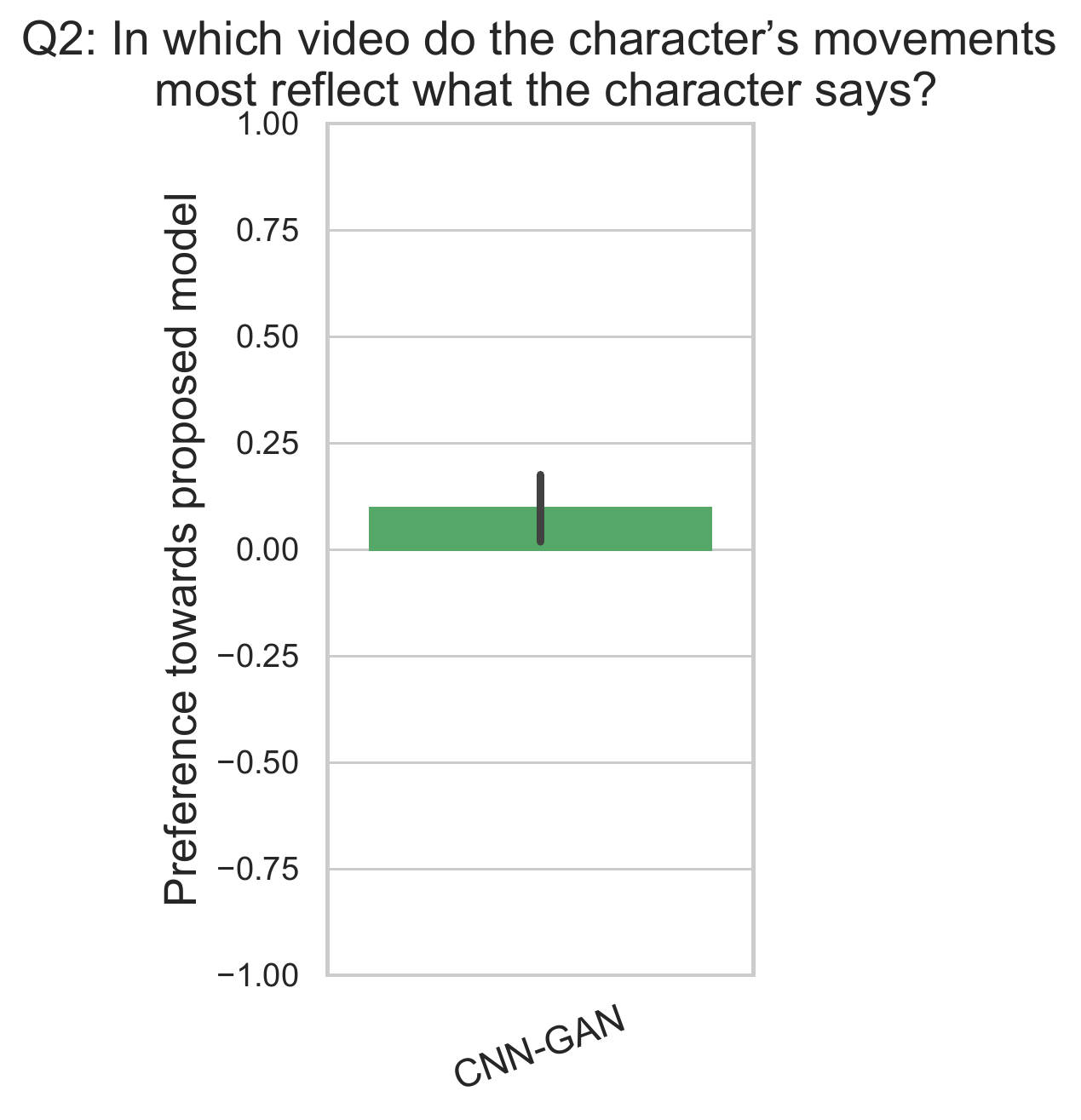}}
\hfill
\caption{\new{Results of Perceptual Study 2: comparing with the state-of-the-art in pairwise preference tests. The bars show the preference towards the full model with 95\% confidence intervals.}}
\label{fig:base_study}
\vspace{-3mm}
\end{figure}

Like for Perceptual Study 1, we analyzed the responses using binomial tests excluding ties followed by Holm-Bonferroni correction.
The results are shown in Figure~\ref{fig:base_study}. Our model was preferred over the CNN-GAN baseline for Q1 with $p$\textless.0001 and for Q2 with $p$\textless.02, indicating that the gestures generated by our model were perceived as more human-like and better reflected what the character said.

}

\subsection{Comparison with the Ground Truth}
\new{We also} compared \new{our model} to the ground-truth gestures using the same procedure as before.
In this study (Perceptual Study 3), 20 participants ($\mu$ age = $39.1\pm8.4$; 9 male, 11 female) remained after excluding 31 participants through the same criteria as in Perceptual Study 1. 
$N$ = 18 were from the USA. 
There was a very substantial preference for the ground-truth motion (between 84 and 93\%) across all questions.
All differences were statistically significant according to Holm-Bonferroni-corrected binomial tests ignoring ties.

\new{
\subsection{What Do ``Semantic'' Gestures Even Mean?}

Finally, we evaluated if using text input helps our model to produce more semantically-linked gestures, such as iconic, metaphoric and diectic. To this end, we compared our best model (No PCA) with and without text information in the input: the first variant of this model received both audio and text, while the second one received only audio as input. 

We asked three annotators
to select which segments out of 50 test segments for both conditions that were semantically linked with the speech content. The annotators were all male and had an average age of 25.3 years. They were not aware of our research questions.

The results of this annotation were interesting and surprising: while all of them marked more gestures to be semantically linked with the speech content for the model that used text than the model without text (2 vs 0, 21 vs 9 and 9 vs 4), they had very low agreement: Cronbach's alpha was below 0.5. The low agreement on which segment were semantic indicates that it is very subjective which gestures should be classified as semantically linked, which makes this and any similar evaluation challenging.


}

\section{Conclusions and future work}
\label{sec:concl}
We have presented a new machine learning-based model
for co-speech gesture generation.
To the best of our knowledge, this is the first data-driven model capable of generating continuous gestures linked to both the audio and the semantics of the speech. 

\new{We evaluated different architecture choices and compared our model to an audio-based state-of-the-art baseline using both objective and subjective measures. All the study materials are publicly available at \href{https://figshare.com/projects/Gesticulator/87128}{figshare.com/projects/Gesticulator/87128}.} Our findings indicate that:


\begin{enumerate}
    \item Using both modalities of the speech -- audio and text -- can improve \new{continuous} gesture-generation models.
    \item Autoregressive connections, while not commonplace in contemporary gesture-generation models, can enforce continuity of the gestures, without vanishing-gradient issues and with few parameters to learn. We also described a training scheme that prevents autoregressive information from overpowering other inputs.
    \item PCA applied to the motion space (as used in \cite{yoon2018robots}) can restrict the model by removing perceptually-important variation from the data, which may reduce the range of gestures. 
    \item The gestures from our model were preferred over the CNN-GAN \cite{ginosar2019learning} baseline by the study participants.
\end{enumerate}

\new{The main limitation of our work is that it requires an annotated dataset (with text transcriptions), which is 
labor-intensive. To overcome this, one could consider training the model directly on transcriptions from Automatic Speech Recognition.}

\edit{Additionally, the vocabulary used in this dataset is sub-optimal. As we can see in the frequency table provided at \href{https://preview.tinyurl.com/y22h6rtt }{tinyurl.com/y22h6rtt}, out of the 50k total words there are 4230 unique words and the first 8 words account for 30\% of all words spoken. This makes it challenging to learn semantic relations between gestures and text. }

Future work \new{also} involves making the model stochastic (as in \cite{alexanderson2020style}), \edit{using larger datasets (such as \cite{lee2019talking})} and further improving the semantic coherence of the gestures, for instance by treating different gesture types separately.

\section*{Acknowledgement}
The authors would like to thank Andre Pereira, Federico Baldassarre and Marcus Klasson for helpful discussions. 
This work was partially supported by the Swedish Foundation for Strategic Research Grant No.: RIT15-0107 (EACare), by the Swedish Research Council projects 2017-05189 (CrowdVR) and 2018-05409 (StyleBot) and by the Wallenberg AI, Autonomous Systems and Software Program (WASP) funded by the Knut and Alice Wallenberg Foundation.


\bibliographystyle{ACM_reference_format}
\balance
\bibliography{refs}

\end{document}